\documentstyle[epsfig]{aipproc}

\begin{document}
\title{A Thermal-Nonthermal Inverse Compton Model for Cyg X-1}

\author{A. Crider$^*$, E. P. Liang$^*$, I. A. Smith$^*$\\
D. Lin$^*$ and M. Kusunose$^{\dagger}$}
\address{$^*$Department of Space Physics and Astronomy, Rice University\\
Houston, TX 77005-1892\\
$^{\dagger}$Kwansei University, Uegahara Ichiban-cho\\
Nishinomiya 662, Japan}

%\lefthead{LEFT head}
%\righthead{RIGHT head}
\maketitle

\begin{abstract}
Using Monte Carlo methods to simulate the inverse Compton scattering of 
soft photons, we model the spectrum of the Galactic black hole candidate
Cyg X-1, which shows evidence of a nonthermal tail extending beyond
a few hundred keV.  We assume an \emph{ad hoc} sphere of leptons,
whose energy distribution consists
of a Maxwellian plus a high energy power-law tail, and inject
0.5 keV blackbody photons.  The spectral data is used to constrain
the nonthermal plasma fraction and the power-law index assuming a 
reasonable Maxwellian temperature and Thomson depth.  A small but non-negligible
fraction of nonthermal leptons is needed to explain the power-law tail. 
\end{abstract}

\section*{Introduction}

Cygnus X-1 is the brightest and most studied Galactic black hole candidate.
Its high-energy spectrum has been extensively modeled 
in attempts to determine what processes produce the continuum emission.
While several models debate the detailed geometry of the system 
[1-4],
many agree that the 10-200 keV portion of the spectrum is a result of 
inverse Compton scattering of soft X-ray photons.
Early modeling attempts assumed a single temperature Maxwellian
plasma [5].
However such simple models do not adequately fit the observations
[6,7].  More complex models, such as one with two
plasma regions of different temperature, increase the goodness-of-fit
substantially [3].

When Cyg X-1 is in the normal ($\gamma_{2}$)
state, a high-energy power-law is observed in the $\gamma$-ray continuum
[8].
This is typically attributed to some nonthermal process. 
It is possible that the power-law is due to inverse Compton
scattering off of a high-energy power-law tail in the plasma energy
distribution.  If so, both the emission above and below
300 keV could originate from the same plasma.

\section*{Fitting the Cyg X-1 Spectrum}

We wish to see if the high-energy spectrum of Cyg X-1 can be explained simply
by Comptonization from a single plasma region with a Maxwellian+power-law
energy distribution.
We simulated Comptonized spectra with a Monte Carlo code
based on the algorithm of Pozdnyakov, Sobol, and Syunyaev [9].  We compared
the results to combined BATSE and COMPTEL data from 1991
[8,10].  Though this data
has already been unfolded through the detector response
using an assumed model, it serves to give us an approximate solution.

The parameters that define the shapes of our simulated spectra are 
the temperature of the thermal leptons (electrons and pairs) $\rm{T_{e}}$,
the Thomson depth $\tau_{\rm{T}}$, 
the fraction of nonthermal leptons $\xi$, 
and the energy index of the nonthermal leptons $\rm{p}$.
Generating and interpolating over a 4-dimensional parameter space
is computationally intensive and for 
this study we fixed both $\rm{T_{e}}=65~keV$ and $\tau_{\rm{T}}=2.45$. 
A value of $\rm{T_{e}}$ very close to this was determined for this data
assuming single temperature analytic Comptonization [8].
The parameter $\tau_{\rm{T}}$ was determined from the spectral index 
$\alpha$ of the 30 to 70 keV spectra, as suggested by 
Pozdnyakov, Sobol and Syunyaev [9], where the equations  
\begin{equation}
\gamma = \frac{{\pi}^{2}}{3} 
	 \frac{{mc}^{2}}{ {(\tau_{\rm{T}} + \frac{2}{3})}^{2} T_{e}}
\end{equation}
and
\begin{equation}
\alpha = - \frac{3}{2} + \sqrt{\frac{9}{4} + \gamma}
\end{equation}
are solved for $\tau_{\rm{T}}$.
This left two free spectral shape parameters, $\xi$ and $\rm{p}$.  We
calculated a $6 \times 6$ grid of simulated spectra,
with $\xi$ ranging logarithmically from 0.25\% 
to 8.0\% and $\rm{p}$ ranging from 3.25 to 4.5.  These grid points
were chosen based on our experience with this code applied
to other astrophysical phenomena [11,12].
We iterated our code 
until the statistical signal-to-noise within the range of 30 keV to 2020 keV
was less than 10 in each of the 22 bins.  Bins above 500 keV were smoothed
in a manner similar to that in Pozdnyakov, Sobol and Syunyaev [9].  
A simple spline algorithm allowed us make our model continuous
so we could evaluate a $\chi^{2}$ between our simulation 
and the discrete data.
  
\begin{figure}[] % fig 1
\centerline{\epsfig{file=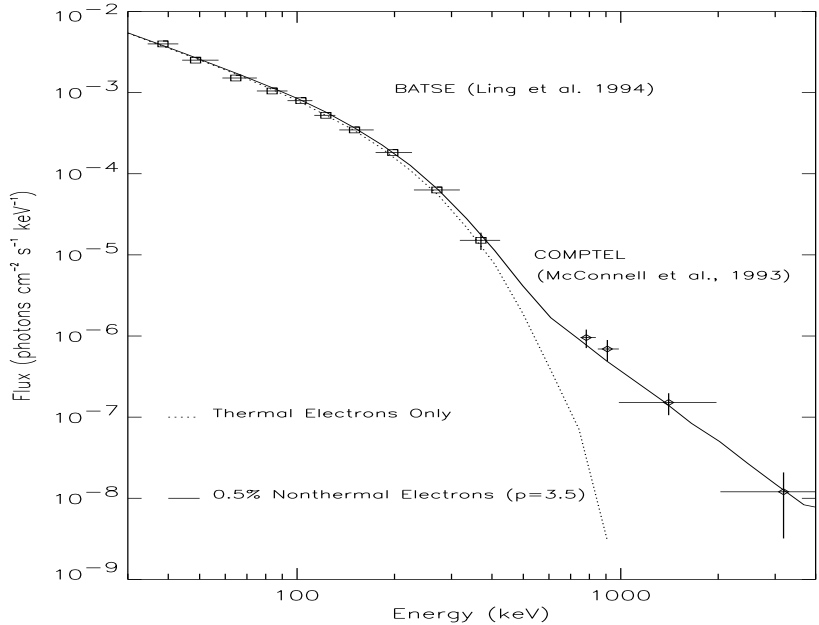,
height=3.5in,width=4.5in}}
\vspace{10pt}
\caption{Fit of Maxwellian+nonthermal Comptonization model to
unfolded spectrum of Cygnus X-1 where
$\rm{T_{e}}=65~keV$, $\tau_{\rm{T}}=2.45$, $\xi=0.5\%$, and $\rm{p}=3.5$. 
Also plotted for comparison is a fit of the same model with no nonthermal
electrons or pairs.}
\label{fig1}
\end{figure}

Figure \ref{fig1} shows a reasonable fit of our code to
the BATSE and COMPTEL data, where $\xi=0.5\%$ and $\rm{p}=3.5$.
To see what range of the parameter space is acceptable, we next examined 
$\chi^{2}$ over the entire grid.  In Figure 2, we plot the confidence levels
of fits on this grid.  Contours are drawn here assuming a 
polynomial interpolation between grid points..  This shows that for a fixed 
$\rm{T_{e}}=65~keV$ and $\tau_{\rm{T}}=2.45$,  there is a 68.3\%
(1$\sigma$) confidence that $\xi$ lies  
between 0.25\% and 1.0\% and $\rm{p}$ is between 3.25 and
4.25.

We remind the reader that this is a crude way 
to test a model since we are comparing to already unfolded data.
However, by showing a reasonable match between our simulated spectra 
and the unfolded spectra, we can be confident that more a rigorous procedure 
would also work. 

\begin{figure}[] % fig 2
\centerline{\epsfig{file=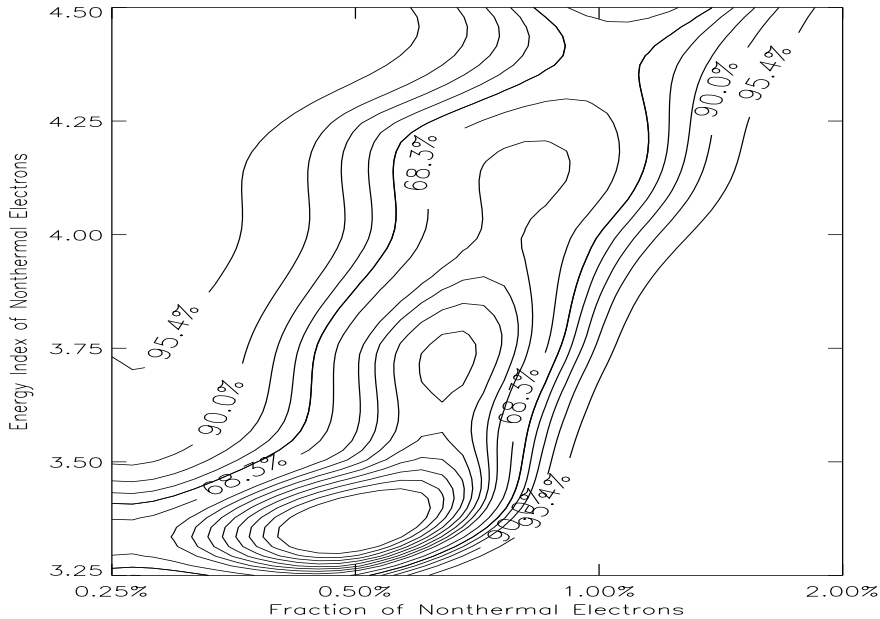,height=3.5in,width=4.5in}}
\vspace{10pt}
\caption{Confidence level contours in $\xi-\rm{p}$ parameter space
for fits of model to Cyg X-1 spectrum. Confidence levels are calculated from
goodness-of-fit statistic $\chi^{2}$ assuming 11 degrees of freedom
(14 data points - 3 fit parameters). See Press et al. 1993 for details.}
\label{fig2}
\end{figure}

\section*{Summary and Caveats}

We find that the X-ray/$\gamma$-ray spectra of the normal state of 
Cygnus X-1 can roughly be reproduced by Comptonization of 0.5 keV 
blackbody photons through a combination thermal-nonthermal plasma.
For $\rm{T_{e}}$=65 keV and 
the Thomson depth $\tau_{\rm{T}}$=2.45, we determine that
the fraction of nonthermal leptons ${\xi=0.5\%}^{+0.5\%}_{-0.25\%}$
and the energy index of the nonthermal leptons $\rm{p}={3.5}^{+0.75}_{-0.25}$.

While these results place first-order limits on the nonthermal lepton 
distribution, several modifications to these procedures would be necessary
in order to reliably determine the allowed parameter space.
The current Monte Carlo code does not require self-consistency between 
the lepton distribution and the radiative cooling.  This obviously must
be corrected to produce physically meaningful results.
It is also necessary 
to generate a four-dimensional grid of simulated spectra to allow all four
shape parameters to vary.  This grid should also extend to lower values of 
$\xi$ and $\rm{p}$ since our current grid does necessarily exclude 
solutions in this portion of the parameter space.
Finally, any future work would require folding our simulated spectra through 
the detector  
responses to allow direct comparison with the $\gamma$-ray count data. 

\section*{Acknowledgements}

Funding for this work is provided through a NASA GSRP Fellowship from 
Marshall Space Flight Center and NASA Grant NAG5-3824.

\end{document}